# Exact Resistance of an Orifice in a 2D Membrane Blocked by a Cylindrical Obstruction

Martin Charron, Vincent Tabard-Cossa*

150 Louis-Pasteur Private, Department of Physics, University of Ottawa, Ottawa K1N 6N5, Canada

*Corresponding Author: tcossa@uottawa.ca

**Abstract**

An exact solution is presented for the resistance of an orifice in a 2D membrane separating two infinitely large conductive reservoirs and obstructed by an infinitely long cylinder. The solution is obtained by constructing a curvilinear coordinate system that captures the symmetry of the obstructed system with constant-coordinate surfaces mapping the system boundaries, and by integrating the resistive contributions of infinitesimally thin equipotential slices. As commonly done when assessing the resistance of fluidic channels of finite length, the exact expression of the obstructed 2D orifice can be used as the access region of obstructed cylindrical channels and will thus find use in single molecule sensing applications.

Resistive pulse sensing is a single-entity technique that allows counting and characterizing particles or molecules suspended in a conductive fluid by measuring the reduction in ionic current, or equivalently the increase in resistance, induced as they pass through a pore or a fluidic channel.[1–6] Since resistive pulse amplitudes are related to the obstructing object's dimensions,[7–15] analysis of the ionic currents in blocked states of the pore allows the extraction of molecular sizes and shapes, as has been demonstrated with the sizing of mammalian cells, nanoparticles, extracellular vesicles, and more recently at the nanoscale with the sizing and shaping of proteins.[16–18] The accuracy of the entity's physical information extracted from resistive pulses is directly correlated to the accuracy of models predicting the ionic response of fluidic channels in their open and blocked states.

The resistive contributions of a cylindrical channel of length $L$ and radius $r_p$ are commonly separated into the pore interior and exterior, the latter of which is commonly referred to as its access regions. The resistance of an open pore $R_o$ is thus treated as resistors in series $R_o = L/\sigma\pi r_p^2 + 1/2\sigma r_p$, where the first term corresponds to the ideal resistance of a cylindrical channel (i.e. the pore interior) of length $L$ and radius $r_p$, and the second term, the access resistance, corresponds to the resistance of an orifice of radius $r_p$ in a 2D insulating membrane, i.e. twice the resistance between a circular disk electrode and an infinitely large hemisphere (Fig. 1a).[19] Similarly, in the presence of a current-obstructing entity spanning both the interior and exterior of a channel, the resistance of a cylindrical pore is calculated as $R_b = R_b^{cyl} + R_b^{2D}$, where $R_b^{cyl}$ and $R_b^{2D}$ are the interior and access region resistance contributions in the blocked state.

Double-stranded DNA (dsDNA) serves as a fundamental analyte for many resistive pulse sensing experiments[20–23] and often acts as a standard 'molecular ruler' for the size characterization of nanoscale channels via conductance measurements.[24,25] During passage through cylindrical nanopore, double stranded DNA is often modeled as an upright infinitely long insulating cylinder due to its high persistence length (~50 nm) compared to its cross-sectional radius ($r_c \sim 1.1$ nm). The interior resistance of the channel is thus trivially found from geometrical arguments, $R_b^{cyl} = L/\sigma\pi(r_p^2 - r_c^2)$. Although different approximate expressions have been published for the access resistance of channels obstructed by insulating cylinders $R_b^{2D}$,[25–27] no exact solution yet exists. In this work, we report an exact analytical expression for the resistance of a 2D pore obstructed by an infinitely long insulating cylinder, $R_b^{2D}$, obtained by introducing a radially offset oblate spheroidal coordinate system whose constant variable surfaces naturally describe the equipotential surfaces and field lines of the system. This exact and closed form solution provides the optimal expression for the access resistance of a channel blocked by DNA or any rigid cylindrical molecules such as viruses[28] or other linear polymers.

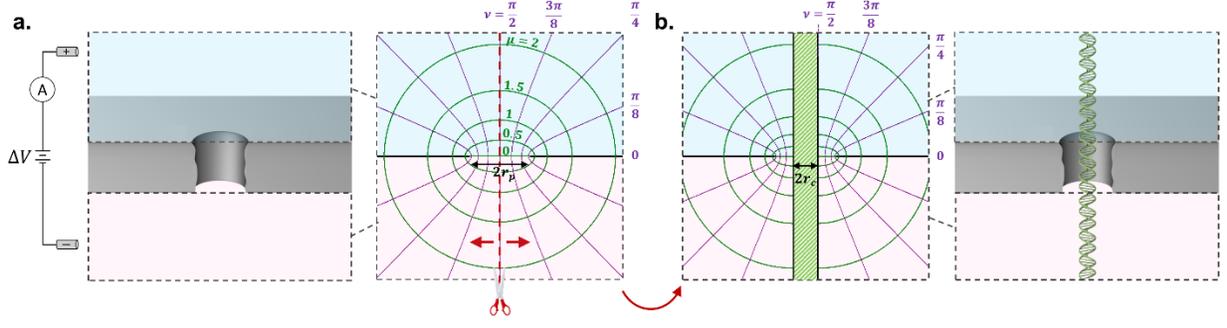

**Figure 1. a)** Oblate spheroidal coordinates are the natural coordinate system for determining the electric response of unobstructed 2D orifice and modeling the access regions of finite-length pores. **b)** Radially offset oblate spheroidal coordinates are constructed to be the natural coordinate system for determining the electric response of 2D pores obstructed by infinitely long insulating cylinders, and modeling access regions in the presence of DNA.

The setup for the model of $R_b^{2D}$ is shown in Figure 1b, wherein a potential difference $\Delta V$ is applied across an orifice of radius $r_p$ in an infinitely thin (2D) membrane blocked by a cylindrical obstruction of radius $r_c$ and of infinite length. Since both reservoirs are assumed to be infinitely large and uniformly conductive, the electrodes are considered as infinitely large hemispheres with potentials of $\pm\Delta V/2$, and the membrane and cylindrical obstruction surfaces are insulating and uncharged. Just like cylindrical coordinates for coaxial electrodes, and spherical coordinates for concentric spherical electrodes, we wish to employ a coordinate system that greatly simplifies electrostatic calculations by capturing the symmetry of the system and its boundary conditions. To this end, as shown in Figure 1a, we first note that the natural coordinate system for an unobstructed 2D orifice is the oblate spheroidal coordinates system $(\mu, \nu, \phi)$.

$$x = r_p \cosh\mu \cos\nu \cos\phi \qquad (1)$$
$$y = r_p \cosh\mu \cos\nu \sin\phi$$
$$z = r_p \sinh\mu \sin\nu.$$

In this coordinate system, constant-$\mu$ surfaces represent oblate spheroid surfaces with $\mu = 0$ corresponding to the circular pore opening and $\mu \to \pm\infty$ to infinitely large hemispheres. Similarly, constant-$\nu$ surfaces correspond to one-sheet hyperboloidal surfaces with $\nu = 0$ corresponding to the membrane surface and $\nu = \pi/2$ to the z axis. Due to the coordinate system mapping the orifice system's boundary conditions, i.e. constant-$\mu$ surfaces mapping to the electrodes and constant-$\nu$ surfaces mapping to the insulating 2D membrane surface, equipotential surfaces correspond to constant-$\mu$ oblate spheroid surfaces, which greatly simplifies electrostatics calculations. Namely, using this coordinate system, the resistance of an unobstructed 2D pore or radius $r_p$ can be shown to be $R_o^{2D} = (2\sigma r_p)^{-1}$.[19]

To find the exact resistance of a 2D orifice obstructed by an insulating cylinder (Fig. 1b), a curvilinear coordinate system is constructed that captures the obstructed system's symmetry and boundary conditions. This is achieved by radially offsetting the oblate spheroidal coordinates (Eq. 1) by $r_c$ from the $z$-axis (red arrows, Fig. 1a), and rescaling the $x$, $y$, and $z$ scales to ensure that the annulus at the pore mouth has a width of $r_p - r_c$, and that orthogonality is preserved:

$$x = \left((r_p - r_c)\cosh\mu \cos\nu + r_c\right)\cos\phi \quad (2)$$
$$y = \left((r_p - r_c)\cosh\mu \cos\nu + r_c\right)\sin\phi$$
$$z = (r_p - r_c)\sinh\mu \sin\nu,$$

where $\mu \in (-\infty, \infty)$, $\nu \in [0, \pi/2]$ and $\phi \in [0, 2\pi]$. In this coordinate system, constant-$\mu$ surfaces correspond to radially offset oblate spheroids:

$$\frac{\left(\sqrt{x^2+y^2}-r_c\right)^2}{(r_p-r_c)^2 \cosh^2\mu} + \frac{z^2}{(r_p-r_c)^2 \sinh^2\mu} = 1 \quad (3)$$

These offset spheroids naturally map the conductive volume by going from an annulus with inner and outer radii of $r_c$ and $r_p$ at $\mu = 0$, to a punctured infinitely large hemisphere at $\mu \to \pm\infty$. Similarly, constant-$\nu$ surfaces are radially offset one-sheet hyperboloids, and map the conductive volume from the insulating 2D membrane surface at $\nu = 0$ to the insulating cylinder surface at $\nu = \pi/2$. Because of the symmetry of this offset coordinate system and the mapping of its boundary conditions, equipotential surfaces naturally correspond to constant-$\mu$ surfaces (Eq. 3) and electric field lines are aligned along constant-$\nu$ surfaces in the $\hat{\mu}$ direction. In this coordinate system, the scaling factors are $h_i = \sqrt{\left(\frac{\partial x}{\partial i}\right)^2 + \left(\frac{\partial y}{\partial i}\right)^2 + \left(\frac{\partial z}{\partial i}\right)^2}$ for $i = \mu, \nu, \phi$ and correspond to:

$$h_\mu = h_\nu = (r_c - r_c)\sqrt{\sinh^2\mu + \sin^2\nu} \quad (4)$$
$$h_\phi = (r_c - r_c)\cosh\mu \cos\nu + r_c.$$

The exact resistance of the obstructed 2D pore can be obtained by partitioning its volume into equipotential slices of infinitesimal thickness $d\mu$ and summing their resistive contributions $dR$, i.e. $R_b^{2D} = \int dR = \int dV/I$, where $dV$ and $I$ are the $\mu$-dependent voltage drop and current across equipotential surfaces.[29] Given that $I = \sigma \int_0^{2\pi}\int_0^{\pi/2} E\, h_\nu h_\phi\, d\nu\, d\phi = \sigma \frac{dV}{d\mu}\int_0^{2\pi}\int_0^{\pi/2} h_\phi\, d\nu\, d\phi$, the exact resistance expression can be found:

$$R_b^{2D} = \frac{1}{\sigma}\int_{-\infty}^{\infty}\frac{d\mu}{\int_0^{2\pi}\int_0^{\frac{\pi}{2}} h_\phi dv\, d\phi}$$

$$R_b^{2D} = \frac{4}{\pi^2 \sigma r_c \sqrt{\frac{4}{\pi^2}\left(\frac{r_p}{r_c}-1\right)^2 - 1}} \tan^{-1}\sqrt{\frac{\frac{2}{\pi}\left(\frac{r_p}{r_c}-1\right)-1}{\frac{2}{\pi}\left(\frac{r_p}{r_c}-1\right)+1}}$$

$$R_b^{2D} = \frac{2}{\pi^2 \sigma r_c \sqrt{1-\frac{4}{\pi^2}\left(\frac{r_p}{r_c}-1\right)^2}} \ln\left[\frac{1+\sqrt{1-\frac{4}{\pi^2}\left(\frac{r_p}{r_c}-1\right)^2}}{\frac{2}{\pi}\left(\frac{r_p}{r_c}-1\right)}\right]. \tag{5}$$

Although arguments of Equation 5 can be complex, both forms are equivalent and purely real for all values of $r_c/r_p$ due to the complex relationship $2i\tan^{-1}(z) = \ln(1-iz)/(1+iz)$.

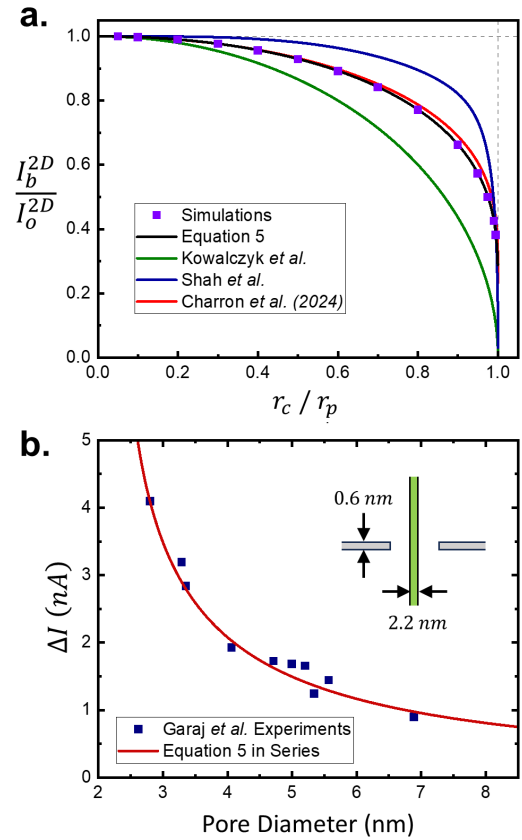

**Figure 2. a)** Plot of equation 5, i.e. normalized conductance, $I_b^{2D}/I_o^{2D} = (R_b^{2D}/R_o^{2D})^{-1}$ versus $r_c/r_p$ (black). Approximations of $G_b^{2D}/G_o^{2D}$ from Kowalczyk et al.[26] (green), Shah et al.[27] (blue), and Charron et al.[25] (2024, red)

are plotted alongside the exact 2D solution for comparison. Values obtained from finite element simulations are plotted, agreeing closely (<0.5% difference) with Equation 5. **b)** Plot of current blockages estimated from Figure 3 of Garaj *et al.*[30] who measured DNA passages in pores of different diameters in graphene membranes. Red line is not a fit, and corresponds to resistive contributions from Equation 5 for the access regions and $L/\sigma\pi(r_p^2 - r_c^2)$ for the pore interior, with a membrane thickness $L$ = 0.6 nm as per the original work. The pore diameters reported correspond to values measured by TEM. Consistent with prior work, 0.2 nm was subtracted from TEM-measured pore diameters and added to B-form of DNA's 2.0 nm bare diameter to account for hydration and condensation effects when calculating resistance.

To visualize Equation 5, Figure 2a plots, in black, the normalized blocked state current $I_b^{2D}/I_o^{2D} = \left(R_b^{2D}/R_o^{2D}\right)^{-1}$ versus the normalized cylindrical radius $r_c/r_p$ alongside approximative equations from Kowalczyk *et al.* (green),[26] Shah *et al.* (blue),[27] and Charron *et al.* (2024, red).[25] Figure 2 also shows finite element simulations for 2D pores obstructed with a cylinder (purple), and were used to validate the accuracy of Equation 5. The modeling of access resistance has been progressively refined over time. While the foundational approximations provided by Kowalczyk *et al.*[26] can diverge from the exact expression by up to 100% under certain conditions, our previous model[25] improved this agreement to within a maximum of 15% deviation. Building upon these efforts, the newly derived, now exact solution introduced here refines this geometric framework even further, exhibiting a near perfect agreement with a negligible discrepancy of under 0.5% with the simulations, which we simply attribute to numerical errors from finite element simulations.

To demonstrate the real-world applicability, Figure 2b compares the predictions of Equation 5 to experimental results from Garaj *et al.*[30] who reported the blockage amplitude for DNA translocations through pores of varying diameters in graphene membranes in 3M KCl solution. As discussed above, resistors in series were used to model the graphene device resistance, $R_b = R_b^{cyl} + R_b^{2D}$, where $R_b^{cyl} = L/\sigma\pi(r_p^2 - r_{DNA}^2)$ and Equation 5 is used for access resistance $R_b^{2D}$. Model predictions were obtained by inserting the experimental values from the original work, $L$ =0.6 nm $\sigma$ =27.5 S/m and $r_{DNA}$ =1.1 nm, into the equations. With no parameters fudging involved, Figure 2b shows excellent agreement between experimental data (blue) and model predictions, demonstrating the quantitative value of this exact access resistance expression for access-dominated experiments.

To gain more intuitive physical insights and a better understanding of its scaling, Equation 5 can be expanded into a series expression around $r_c/r_p = 0$:

$$\frac{R_b^{2D}}{R_o^{2D}} = 1 + \left(\frac{\pi^2}{8} - 1\right)\frac{r_c^2}{r_p^2} + \left(\frac{5\pi^2}{24} - 2\right)\frac{r_c^3}{r_p^3} + \cdots. \tag{6}$$

Equation 6 shows that the blocked-state resistance is simply a product of the open-state resistance of a 2D pore (Eq. 2) and a non-dimensional correction factor dependent only on powers of $r_c/r_p$. Interestingly, the fractional resistive pulse amplitude $\Delta R_b^{2D}/R_o^{2D} = (R_b^{2D} - R_o^{2D})/R_o^{2D}$ scales with the cylindrical cross-sectional area $r_c^2/r_p^2$ in the $r_c \ll r_p$ limit, a scaling which is identical to that expected of very long channels[31] with negligible access regions except for the pre-factor of $(\pi^2/8 - 1) \approx 0.2337$. We recently reported a similar reduced coefficient using a generalized framework for approximating obstructed access resistance.[32] The reason for the reduction of coefficient remains unknown, yet is believed to be a direct result of the non-uniform electric field inside 2D pores, being weakest in its center and approaching infinity near the pore walls,[33] as opposed to the uniform field found in very long channels.

In this work, through the construction of a high symmetry coordinate system, an exact solution was provided for the resistance of a pore in a 2D membrane obstructed by an infinitely long insulating cylindrical object. This expression can be used to model the access resistance of a fluidic channel during the passage of cylindrical molecules such as DNA, viruses or linearized polymers for interpreting resistive pulse sensing results. The application of this model to resistive pulse or nanopore sensing applies best to experimental conditions in which the dominant source of ionic current is from bulk conduction, as opposed to surface conduction, i.e. for uncharged membrane surfaces or in high (> 1M) salt concentrations where surface conduction contributions are minimized. We note that the model is also expected to be less applicable to experiments with very large pores, i.e. $r_p \gg r_c$, where molecules could enter at an angle or shifted from the pore's central axis, as opposed to the vertical and centered position of the cylinder modeled here. Although exact analytical solutions for electrical resistance are known for only a limited number of geometries, the curvilinear coordinate framework presented here yields a rigorously exact expression for an orifice in an ultrathin membrane blocked by a cylindrical polymer. This framework is ready for immediate application in nanopore sensing, which is particularly relevant given the emergence of 2D materials[24,30,34–37] and the ubiquitous role of double-stranded DNA (dsDNA) both as a fundamental carrier in molecular sensing experiments[20–23] and as a standard 'molecular ruler' for characterizing nanoscale channels via conductance measurements.[24,25]

**Acknowledgements**

The authors would like to acknowledge the support of Natural Sciences and Engineering Research Council of Canada (NSERC), [funding reference number RGPIN-2021-04304], and thank Benjamin W. Hyatt for insightful discussions throughout this work.